# Strain-free GaSb quantum dots as single-photon sources in the telecom S-band

*Johannes Michl[1]\*, Giora Peniakov[1], Andreas Pfenning[1], Joonas Hilska[2], Abhiroop Chellu[2], Andreas Bader[1], Mircea Guina[2], Sven Höfling[1], Teemu Hakkarainen[2], Tobias Huber-Loyola[1]*

[1] Julius-Maximilians-Universität Würzburg, Physikalisches Institut, Lehrstuhl für Technische Physik, Am Hubland, 97074 Würzburg, Deutschland

[2] Optoelectronics Research Centre, Physics Unit / Photonics, Faculty of Engineering and Natural Sciences, Tampere University, Finland

## Abstract

Creating single photons in the telecommunication wavelength range from semiconductor quantum dots (QDs) and interfacing them with spins of electrons or holes has been of high interest in recent years, with research mainly focusing on indium based QDs. However, there is not much data on the optical and spin properties of galliumantimonide (GaSb) QDs, despite it being a physically rich system with an indirect to direct bandgap crossover in the telecom wavelength range. Here, we investigate the (quantum-) optical properties of GaSb quantum dots, which are fabricated by filling droplet-etched nanoholes in an aluminum-galliumantimonide (AlGaSb) matrix. We observe photoluminescence (PL) features from isolated and highly symmetric QDs that exhibit narrow linewidth in the telecom S-band and show an excitonic fine structure splitting of $\Delta E_{FSS} = (12.0 \pm 0.5)\,\mu eV$. Moreover, we perform time-resolved measurements of the decay characteristics of an exciton and measure the second-order photon autocorrelation function of the charge complex to $g^2(0) = 0.16 \pm 0.02$, revealing clear antibunching and thus proving the capability of this material platform to generate non-classical light.

1. Introduction

Non-classical light sources are a major building block in quantum communication applications as well as for photonic quantum computing. Emission wavelengths tailored to the telecommunication range are of particular interest since they offer the least absorption and wave packet dispersion of photons in optical fibers[1]. This aspect is crucial for exploiting existing fiber networks by implementing protocols for secure long-range communication, as has been shown for several hundred kilometers of optical fiber[2]. Among several other physical systems like vacancy centers in diamond[3] and trapped atoms[4], semiconductor quantum dots (QDs) offer superior optical properties like high single photon purity and indistinguishability [5]–[9]. Over the last few years, several QD material platforms like indium(gallium)arsenide (In(Ga)As)/ galliumarsenide (GaAs) and indiumarsenide (InAs)/ indiumphosphide (InP)[10] have emerged as resources for non-classical light and spin-photon interfaces in the telecom wavelength range, demonstrating a first two-photon interference experiment over a 2 km fiber channel[11]. This shows the possibility of the platform to scale to measurement device independent quantum key distribution (MDI-QKD) and to extend to quantum repeaters. Especially, the use of a metamorphic buffer (MMB) layer can be utilized to shift the emission of InAs quantum dots, which are well understood spin-photon interfaces[12] with outstanding performance in the NIR wavelength range around 900 nm, into the telecom range and to create polarization-entangled photons[13]. First attempts to increase collection efficiency by integrating telecom InAs QDs into photonic structures in free space[14],[15] as well as fiber-coupled[16] have been demonstrated, and first steps into plug&play devices for quantum key distribution (QKD) in the telecom O-band have been realized[17].

---

\*Email: johannes.michl@physik.uni-wuerzburg.de

While all these approaches cover the need for low losses on the photonic side, the other pressing requirement for memory-based quantum repeaters is control over the matter qubit, which is the spin in a quantum dot. In previous works on InAs/GaAs QDs, we have shown the demonstration of full coherent control of a spin-qubit emitting in the telecom C-band[18]. However, the spin coherence times were very short ($T_2^* = 240$ ps [18]), limiting the useability of the system when approaching implementation of the QDs in quantum repeater applications. The times to which the coherence of an optically active spin-qubit can be extended (e.g. through dynamical decoupling techniques) are, amongst others, governed by the nuclear spins[19]. In this indium plays a dominant role as was experimentally shown in recent years[20], as it has a large nuclear spin of $s_{N,In} = \frac{9}{2}$ (In comparison: $s_{N,As} = \frac{3}{2}, s_{N,Ga} = \frac{3}{2}, s_{N,Sb} = \frac{5}{2}$). An example of an indium-free material platform is GaAs QDs, grown by droplet etching epitaxy on aluminum-gallium- arsenide (AlGaAs), which has been shown to be a source of non-classical light of high quality[21] in the wavelength range of 780 nm. Using GaAs QDs and exploiting decoupling techniques, recent work could extend electron spin coherence times into the millisecond range[22]. Indium-free QDs are thus very promising to extend the coherence times of semiconductor QDs, but the GaAs material system is limited to the short wavelength range. GaSb QDs are a promising indium-free QD candidate to make comparably long coherence times also available in the telecom wavelength range.

In this work, we investigate the unexplored quantum-optical properties of GaSb quantum dots, grown by local droplet etching of nanoholes and subsequent filling. We show first results of time-resolved as well as autocorrelation measurements of emission features from single GaSb QDs emitting in the telecom S-band. Our results pave the way to exploit this material platform, that combines the low losses of photons in the telecom range with possibly long spin coherence times due to the lack of indium.

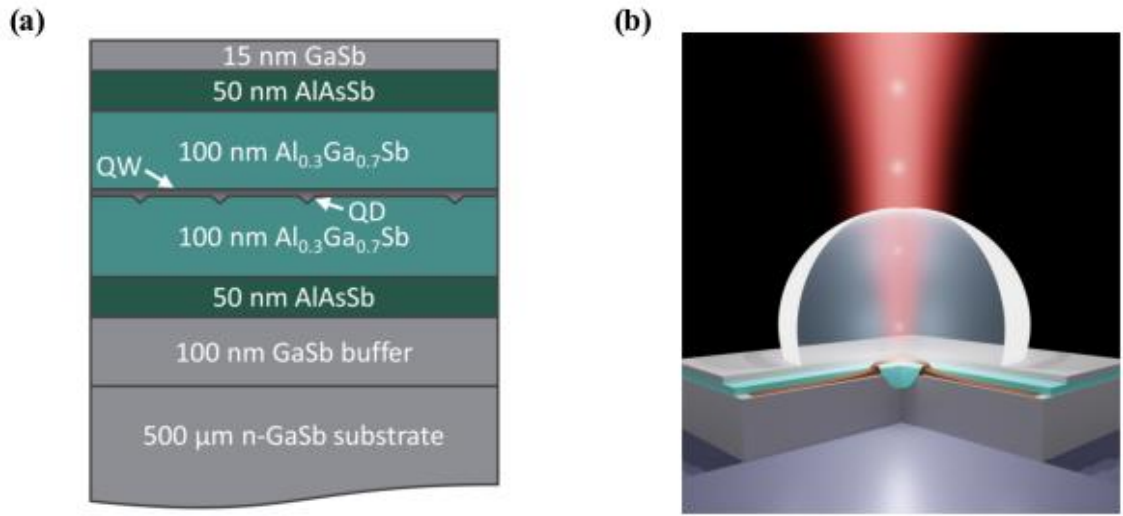

**Figure 1**. Sample design and layer structure. **(a)** Detailed schematic of the layer sequence. A GaSb substrate and buffer is used, followed by 50 nm of AlAsSb. The droplets are etching holes into a 100 nm layer of $Al_{0.3}Ga_{0.7}Sb$ which are then filled with GaSb, followed by an $Al_{0.3}Ga_{0.7}Sb$ layer of identical thickness as the first. The layer structure ends with $50\ nm$ of AlAsSb and a 15 nm GaSb cap. **(b)** Simplified, artistic view of the sample, including the filled nanohole forming a quantum dot, surrounded by the quantum well (QW) area. A solid immersion lens is placed over the whole sample, increasing the collection efficiency of the emitted photons (depicted as bright light bundles). The laser is drawn as a red, gaussian shaped beam.

2. Sample Design and Heterostructure Growth

The GaSb QDs investigated in this study were grown by solid-source molecular beam epitaxy (MBE) on GaSb(100) substrates by GaSb filling of nanoholes etched into AlGaSb via local droplet etching[23]. This method was first developed for the GaAs/AlGaAs system[24],[25] and has been recently extended to the GaSb/AlGaSb system (see previously published articles[26],[27], including more details on the growth of our sample).

The exact heterostructure is provided in Figure 1 (a). After oxide desorption, the growth process is started by depositing a 100 nm thick GaSb buffer followed by 50 nm of aluminarsenideantimonide (AlAsSb) and 100 nm of aluminumgalliumantimonide ($Al_{0.3}Ga_{0.7}Sb$). The local droplet etching is initiated by depositing 1.5 monolayers (MLs) of Al with a 0.3 ML/s flux at a temperature 395 °C under a low Sb-flux of 0.07 ML/s. The excess Al forms a uniform distribution of metal droplets on the surface with a density of $2.6 \times 10^7\ cm^{-2}$, which then etch into the underlying $Al_{0.3}Ga_{0.7}Sb$ matrix forming ~12 nm deep nanoholes[27]. The nanoholes are subsequently filled by GaSb and then encapsulated by $Al_{0.3}Ga_{0.7}Sb$. In this study, a thickness of the GaSb-filling layer of 20 ML was used. To promote carrier capture into the QDs, the $Al_{0.3}Ga_{0.7}Sb$ /GaSb-QD/ $Al_{0.3}Ga_{0.7}Sb$ structure was also encased with $AlAs_{0.08}Sb_{0.92}$. The topmost $AlAs_{0.08}Sb_{0.92}$ is protected from atmospheric oxidation by a thin 15 nm GaSb capping layer. For more efficient collection of the emitted PL, a solid immersion lens (SIL) was placed over the whole sample, as can be seen in the artistic drawing of our sample in Figure 1 (b).

3. Experimental Data and Discussion

3.1 Setup and Methods

The sample was mounted in a closed-cycle cryostat, which allows for keeping a stable position for up to several weeks while keeping the sample at a base temperature of T=1.57 K. For excitation, a fiber coupled continuous-wave laser with a wavelength of 940 nm was used (further called cw-laser) as well as a ps pulsed optical parametric oscillator (OPO) with identical wavelength as laser the cw-laser and a repetition rate of 80 MHz (further called pulsed laser).

In the detection path, polarization optics (quarter-wave plate, half-wave plate, and linear polarizer, further called QWP, HWP, and LP) were installed, to perform polarization-resolved PL-measurements. The QD signal was then fiber coupled and analyzed in a 750 mm Czerny-Turner spectrometer for its spectral components and detected with an indiumgalliumarsenide (InGaAs)-camera. A fiber-coupled variable bandpass filter was used to isolate single emission lines from individual charge complexes for further analysis. To resolve the signal temporally, the filtered signal was redirected onto superconducting nanowire single photon detectors (SNSPDs), connected to correlation electronics, which can be triggered by the laser.

3.2. µ-PL Spectra and Characterization

The fundamental optical properties of individual GaSb QDs are studied by means of micro photoluminescence (µ-PL). The average QD density of $2.6 \times 10^7$ cm$^{-2}$ ensures that individual QDs can be addressed and excited. In agreement with earlier reports[27], we could observe PL-emission in the telecom S-band around 1470 nm.

Figure 2 (a) depicts a spectrum that was obtained by exciting the QD with a wavelength of 940 nm, an excitation power of $P = 1.89$ µW and an integration time of 100 s. The most prominent emission feature (referred to as *main peak* from now on), is located at approximately 1469.2 nm and highlighted by the red background in the spectrum. Surrounding it are various other discrete recombination features at higher wavelengths. These characteristics look similar to the PL from GaAs QDs, grown via droplet etching, under non-resonant excitation[28]. To analyze the underlying dynamics between the emission features and to identify the nature of the studied charge complex, we performed a power series. For this, we increased the laser pump-power from 0.27 µW to 3.84 µW and measured the PL emission spectra. Every thus acquired spectrum was fitted with a Lorentzian multipeak fit to track the evolution of the peaks. The inset in the upper right corner of Figure 2 (a) depicts the integrated intensity of the main peak, extracted from the fits, as a function of the excitation power on the double-logarithmic scale. A linear fit to the logarithmic axes yields a power-law coefficient of 1.03, pointing towards emission from a neutral exciton or a trion. For $P > 2.5$ µW, the main peak PL-intensity saturates.

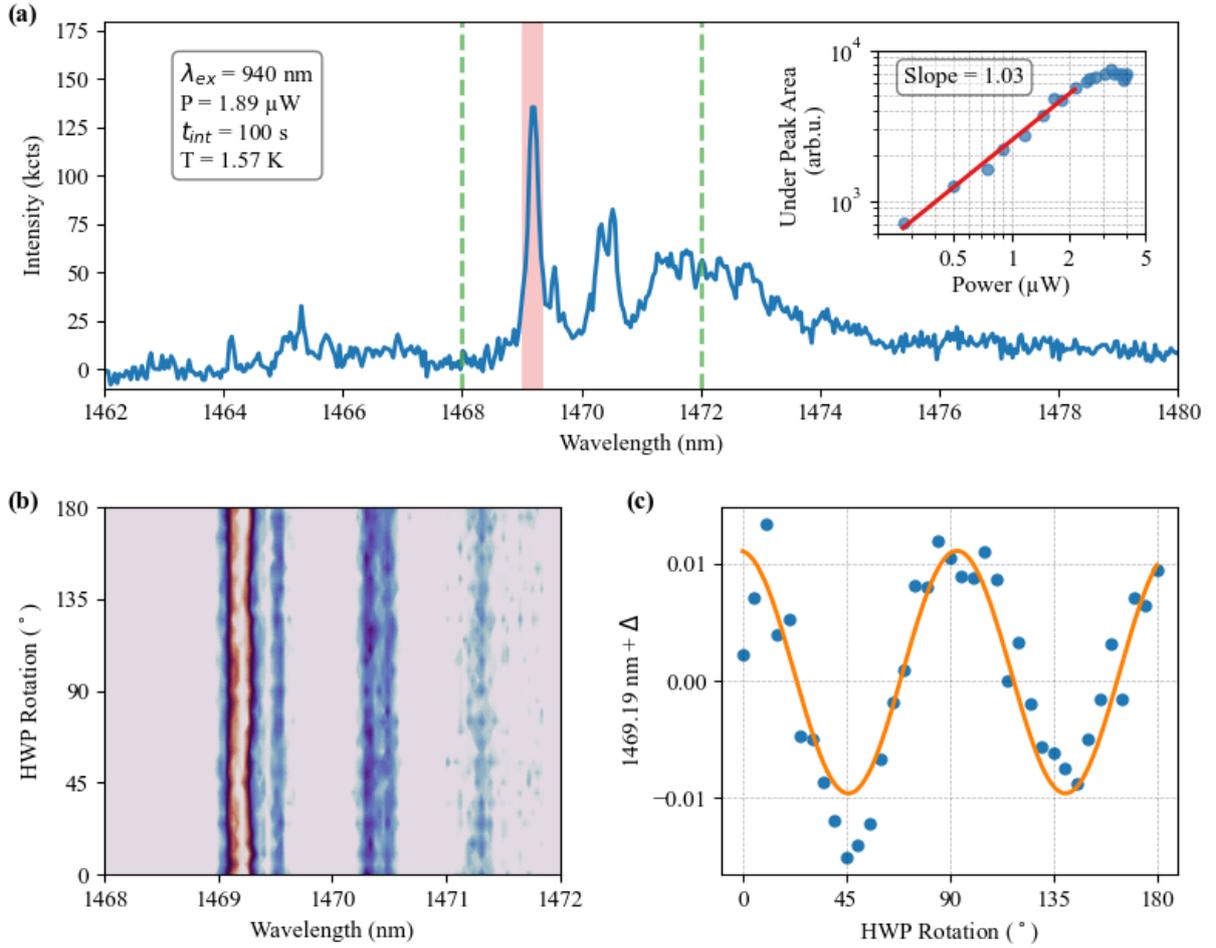

**Figure 2.** Micro photoluminescence for QD charge complex characterization. **(a)** PL spectrum, acquired at $T = 1.57\,\text{K}$ with 100 s integration time under excitation with 940 nm. The brightest line is observed at around 1469.2 nm (main peak) and highlighted with a red square background in the plot. The green dashed lines indicate the wavelength range resolved in (b). **(Inset)** Peak intensity of the main peak, vs. excitation power. Plotted on the double logarithmic scale, a power-law coefficient of 1.03 can be derived from a linear fit function (solid red line). The PL intensity saturates for powers $> 2.5\,\mu W$. **(b)** False color plot of the normalized PL-spectra of a polarization series with the y-axis displaying the rotation of the HWP from 0° to 180°. A slight oscillation of the main peaks center wavelength position with the HWP rotation angle is visible, which is further quantified in **(c)**. Here, the position of the main peak is extracted from a series of fits and plotted versus the HWP rotation. A sinusoidal fit to the data yields a fine structure splitting of $\Delta\lambda_{FSS} = (0.021 \pm 0.001)\,\text{nm}$ or $\Delta E_{FSS} = (12.0 \pm 0.5)\,\mu eV$ respectively. The data displayed in this figure suggest, that the investigated charge complex is an exciton with small FSS.

Further information about the nature of the observed charge complexes was gathered by performing a polarization series. A structural asymmetry of the QD leads to a mixing of the spin states of the exciton and results in a fine-structure splitting (FSS)[29], while a trion on the other hand would not show any dependency on the polarization angle[29]. For this measurement, we placed a HWP in front of a LP in the detection path and rotated it from 0 ° to 180 °, to measure the polarization-dependent spectra in all rectilinear projections. The normalized spectra are displayed in Figure 2 (b) as a function of wavelength and HWP rotation angle in a colored contour plot. The main peak displays a subtle, visible oscillation with the rotation angle that can be further quantified by fitting the series, as is depicted in Figure 2 **(c)**. A sinusoidal fit to the peak position versus the rotation angle yields a splitting amplitude of $\Delta\lambda_{FSS} = (0.021 \pm 0.001)$ nm corresponding to $\Delta E_{FSS} = (12.0 \pm 0.5)$ μeV respectively. Together with the above analyzed power dependency of the peak intensity this suggests that the investigated charge complex is an exciton with a small fine structure splitting (FSS), which can be attributed to the slight asymmetry of the QDs.

3.3. Quantum-Optical Properties

To investigate the quantum-optical properties of the identified exciton, we performed a series of time-resolved measurements. We filtered the spectral line of the main peak using a variable bandpass filter and measured with a temporal resolution of about 20 ps using SNSPDs and correlation electronics.

*3.3.1. Time resolved photoluminescence*

By analyzing the temporal evolution of the PL-intensity measured on the SNSPDs, insight can be gained into the fundamental dynamics of charge carriers in our sample. For this, the pulsed laser was synchronizing the correlation electronics and an excitation wavelength of 940 nm. A pulse duration of ~2 ps were used to excite the QD. Figure 3 (a) displays the time resolved PL data, with the spectrally resolved filtered main peak as an inset in the upper right corner. One can see clearly that there is a filling process of significant time.

Under above-bandgap excitation, multiple charge injection and decay channels are bound to contribute to the observed temporal signature, forming a complex system of coupled and time dependent populations. To account for this theoretically, we assume the model that is illustrated in Figure 3 (b), where at a time $t_0$ after the laser pulse there are three reservoirs (QD, QW Γ-valley and QW $L$-valley ) with three initial populations ($x_0$, $G_0$ and $L_0$) and time dependent evolutions ($X(t)$, $G(t)$ and $L(t)$) respectively. The $L$-valley of the QW has a higher density of states and its feeding into the Γ-valley can be seen in independent measurements of the QW emission (not shown here). Equation (1) formulates the set of coupled differential equations which we use to describe the dynamics in this system:

$$\begin{cases} \dot{X}(t) = -a\,X(t) + b\,G(t) \\ \dot{G}(t) = -(b+c)\,G(t) + d\,L(t), \\ \dot{L}(t) = -d\,L(t) \end{cases} \quad \begin{cases} X(0) = x_0 \\ G(0) = G_0 \\ L(0) = L_0 \end{cases} \quad . \quad (1)$$

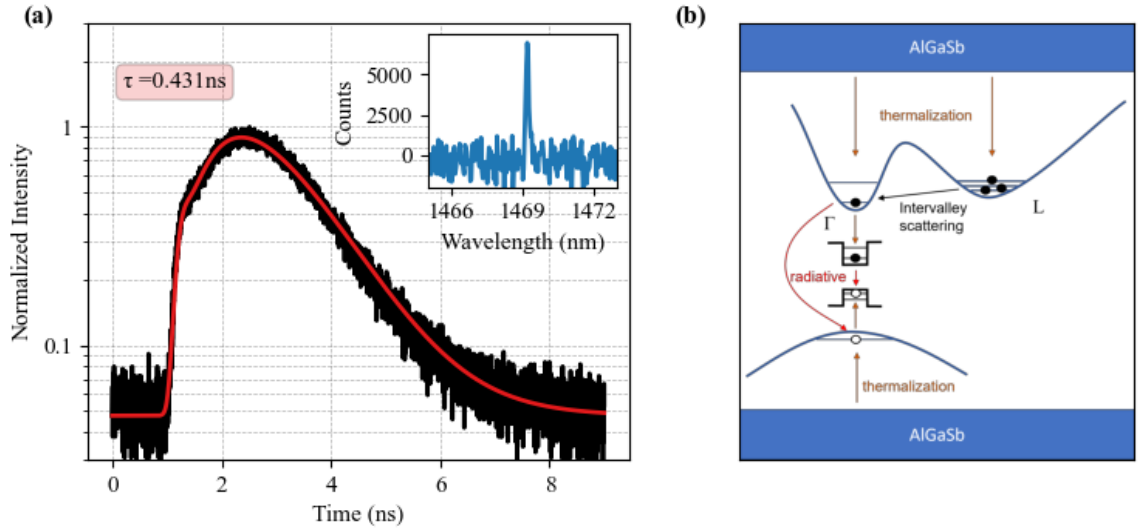

**Figure 3.** Temporal evolution of charge injection into and decay of excitons in the QD. **(a)** Time resolved PL-intensity of the filtered emission line, which is displayed in the upper right inset (10 s integration time, on spectrometer). The data is fitted with the function $X(t)$ as described in Equation (S1). From the fit (solid red line), the QD radiative lifetime can be determined to $T_1^X = (431 \pm 10)$ ps. The overall shape of the time trace is dominated by the long decay time of $4.35\ ns$ from the QW to the QD. **(b)** Theoretical model of the contributing processes. Rates, that must be considered for solving the differential equation describing the observed time trace, are: (i.) radiative decay of charge carriers in the QD excited state to the QD ground state, (ii.) filling of the QD through relaxation from the QW $\Gamma$-valley, (iii.) direct radiative decay from the QW $\Gamma$-valley to the ground state, and (iv.) scattering from the QW $L$-valley to the $\Gamma$-valley. Together with initial populations for QD, $\Gamma$-valley and $L$-valley, these rates form the system of coupled differential equations described in Equation (1).

The temporal evolution of the QD charge carrier population is governed by the decay rate "$a$" of excitons and the filling rate "$b$" from the QW Γ-valley. Meanwhile, the population in the Γ-valley is diminished through this decay to the QD, but also through radiative decay directly to the ground state with rate "$c$". The QW $L$-valley fills the Γ-valley with a rate "$d$". Considering the above given rate equations, the experimentally acquired data can be described by the time dependent evolution $X(t)$. The set of differential equations is therefore solved (cf. supplementary material, equation S1) and fitted to the data as a solid red line in Figure 3 (a). Since our rate equation starts with a given filling at time zero and not with a temporal finite excitation pulse, an error function with width σ is multiplied to $X(t)$. Our true zero where the QD is filled and starts to decay is at about 1 ns in Figure 3 (b). From the fitted parameter "$a$", the radiative lifetime $T_1^X = (431 \pm 10)$ ps can be inferred. A simple exponential fit, which would not follow our rate equation, would give a lifetime of $(930 \pm 9)$ ps, when fitting only times larger $t > 4$ ns.

From the rate equation, the decay time of the quantum well to the quantum dot can be determined to 4.35 ns and is therefore the dominating process for the overall shape of the time trace. Whether the filling rates from this model with many free parameters are actually reflecting the reality should be checked in future independent measurements of the QW decay.

Our time resolved investigations of the excitonic emission line show that under above-bandgap excitation of a reservoir of charge carriers, the complex filling processes from the $L$- and $\Gamma$-valley of the QW to the quantum dot play a predominant role. While these dynamics are of fundamental interest, they also show the need to excite the QDs resonantly in further experiments, as this will eliminate the filling through the QW. Under resonant pumping, an exponential fit of the time trace of the exciton would directly yield its radiative lifetime. Furthermore, delayed re-excitation during one laser pulse from a reservoir could be avoided, which strongly affects the single photon characteristics as we show in the following paragraph.

*3.3.1. Second Order Autocorrelation Measurement*
To evaluate how well single photons, originating from the discrete energy levels inside the QD, can be separated from any unwanted multiphoton background, we performed second order photon autocorrelation measurements. The same excitation conditions as for the measurement of the excitonic time trace were used. In this case, however, the signal was channeled through a 50:50 fiber beam-splitter onto two different channels of the single photon detectors that were connected to the correlation electronics. The correlation electronics were not synchronized to the laser any longer but operated in a "start/stop" manner, where one of the SNSPDs started the correlation measurement, while the other stopped it.

Figure 4 displays the autocorrelation measurement as a histogram of background corrected coincidences versus time delays $\Delta t$ between detection events on the two detectors. The autocorrelation measurement, which is an approximation to the second order correlation function $g^{(2)}(\Delta t)$, shows a clear antibunching behavior of photons around zero time delay. To quantify the value of $g^{(2)}$ at zero time delay, the counts accumulated in the zero peak are normalized on the average of the accumulated counts in each side peak. The coincidence window, i.e., the width of the integration, is set arbitrarily such that most of the counts are included, which in our case results in $8\ ns$. This yields a value $g^{(2)}_{corrected}(0) = 0.16 \pm 0.02$, which is about half the value from the raw data ($g^{(2)}_{raw}(0) = 0.32 \pm 0.01$). However, we want to describe the source as precisely as possible, and our detectors have a significant dark count rate compared to the rate of detected photons. Thus, we believe the background corrected value, where the background is calculated from independently measured dark-/ signal count rates reflects the source properties better. Details on the calculation of the background level as well as the data without background correction (Figure S1) can be found in the supplemental information.

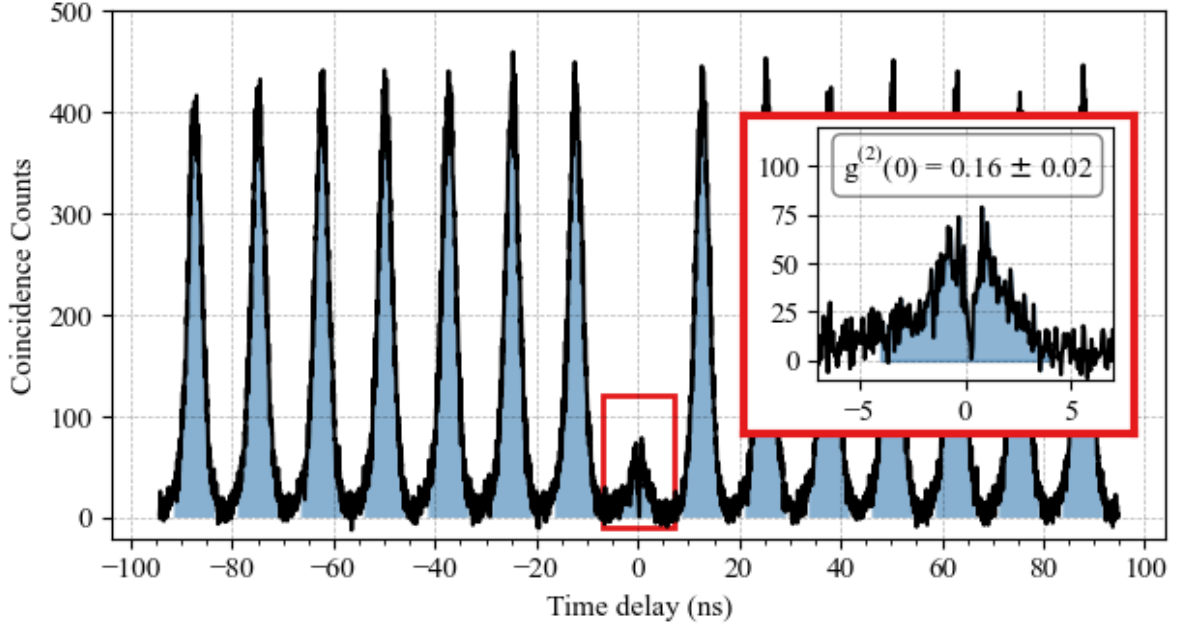

**Figure 4.** Autocorrelation measurement of the exciton recombination photons in Hanbury Brown-Twiss configuration. For zero time delay $\Delta t$, the number of measured coincidences is significantly lower than for non-zero delay times. Comparing the averaged integrated counts of peaks for $t > 0$ s with the peak at $t_0 = 0$ s using equal integration widths of 8 ns, one can derive a $g^2(0)$ value of $0.16 \pm 0.02$. Upper right inset: Magnified depiction of the observed dip in the central peak region, indicated by the red squares.

The inset in the upper right corner of Figure 4 zooms into the central peak, displaying an additional dip in the coincidence counts at $\Delta t = 0$ that reaches down near zero counts. We believe that the counts close to zero time delay accumulated in the central peak, which contribute mainly to the non-zero $g^2(0)$ value, are a direct result of the refilling from the QW, as explained in the lifetime measurement (see section 3.3.1). This leads to multiple excitations during one laser pulse. While this increases the number of photons contaminating the single-photon characteristics of our non-resonantly excited source, it demonstrates the high potential of GaSb QDs to be good single photon sources when exciting the QD resonantly.

4. Conclusion

We have investigated the characteristics of narrow-linewidth PL-features from GaSb QDs, analyzing time resolved PL and the second order correlation function of a stream of photons from an exciton that are the first measurements of their kind on this material platform.
Exciting above-bandgap, the fine structure splitting of the exciton could be determined to $\Delta E_{FSS} = (12.0 \pm 0.5)$ µeV. Moreover, the time-resolved signature of the exciton recombination photons reveals a slow rise of the PL intensity after excitation. This slow rise points towards the existence of multiple charge injection channels. By solving a system of three coupled differential equations, we have been able to fit the data, extracting the radiative lifetime of the exciton to be $T_1^X = (431 \pm 10)$ ps. The complex filling mechanisms from the QW Γ- and $L$-valley when exciting above bandgap also reflect in the autocorrelation measurement, where double excitations during one laser pulse could be observed.

Nevertheless, we have been able to prove the non-classical nature of the PL of the investigated charge complex by measuring a $g^{(2)}(0)$ value of $0.16 \pm 0.02$. A drop of correlated count-events at zero delay time suggests that GaSb QDs have a high potential to be good sources of single photons under resonant pumping.

## Acknowledgements

The University of Wuerzburg group is grateful for financial support by the German Ministry for Research and Education (BMBF) through the project QR.X (FKZ: 16KISQ010) and the State of Bavaria. THL acknowledges funding from the BMBF through the Quantum Futur (FKZ: 13N16272) initiative. The tampere group acknowledges financial support from the Academy of Finland project QauntSi (decision No 323989) and the Business Finland co-innovation project QuTI.(41739/31/2020).

## Author declarations

### Conflict of interest

The authors have no conflicts to disclose.

### Data availability

The data that support the findings of this study are available from the corresponding author upon reasonable request.

# Supporting Information

Raw data of autocorrelation measurement without background correction

Without taking into account the constant, purely statistical contribution of dark counts on the detectors, the histogram of the of the coincidence counts is added with a constant offset, as can be seen in Figure S1. Calculating $g^{(2)}(0)$ with the same intervals as in the main text yields a value of $g^{(2)}_{raw}(0) = 0.32 \pm 0.01$, which is still significantly below 0.5.

Background correction:
To take the statistical contribution of dark counts into account, we measured the dark count rate $r_{1,(2)-\text{dark}}$ of our detector system for 10 s and calculated the expected background contribution together with the measured rate at each detector $(r_1, r_2)$ as:

$$\text{counts}_{BG} = (r_1 \cdot r_{2-\text{dark}} + r_2 \cdot r_{1-\text{dark}}) \cdot t,$$

where $t$ is the total measurement time.

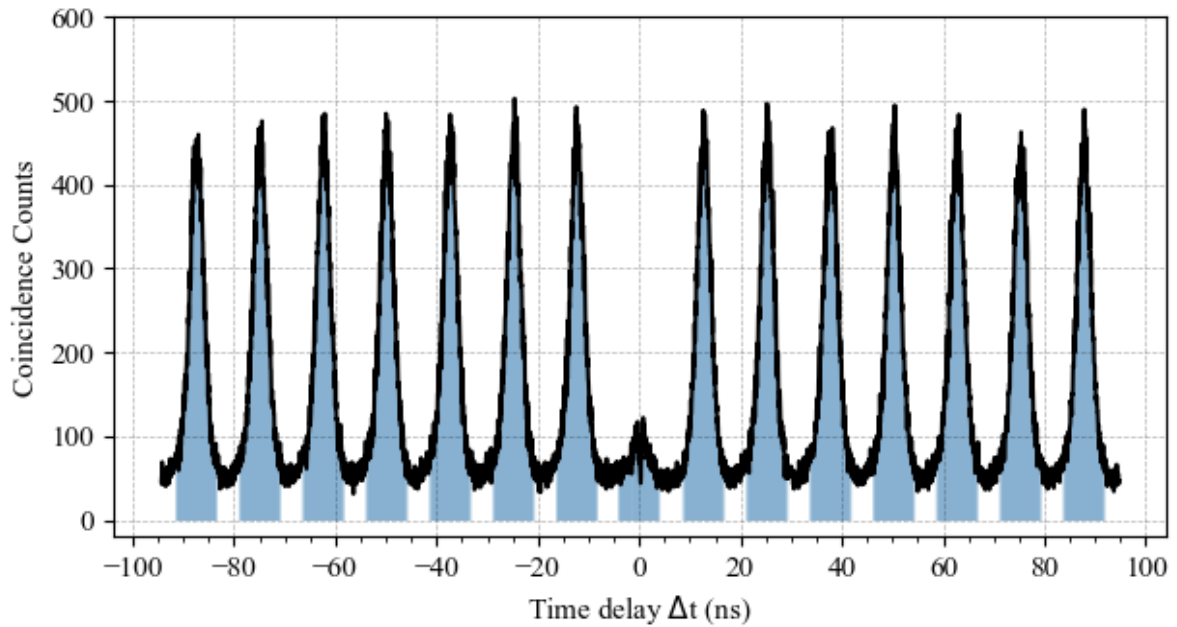

**Figure S1.** Autocorrelation measurement without background subtraction. The uncorrected $g^{(2)}(0)$ value is $g^{(2)}_{raw}(0) = 0.32 \pm 0.01$.

Solution of the system of three coupled differential rate equations

The system of coupled differential rate equations as described in Equation (1) can be solved to the individual components $X(t), G(t)$ and $L(t)$. The solution for $X(t)$, which corresponds to the observable QD radiative decay, is:

$$X(t) = e^{-ct-bt-at}\left(\frac{e^{ct+bt}\left(\left((c+b-a)\,d-a\,c-a\,b+a^2\right)x_0+b\,d\,L_0+(b\,d-a\,b)\,G_0\right)}{(c+b-a)\,d-a\,c-a\,b+a^2} - \frac{(b\,d\,L_0+(b\,d-b\,c-b^2)\,G_0)\,e^{at}}{(c+b-a)\,d-c^2+(a-2b)\,c-b^2+a\,b}\right) + \frac{b\,d\,L_0}{(d^2+(-c-b-a)\,d+a\,c+a\,b)\,e^{dt}}.$$

(Equation S1)

This function was fitted to the dataset that is illustrated in Figure 3 (b) of section 3.3.1.